\documentclass[twocolumn,resetfootnote]{aastex7}

\newcommand\SgrA{Sgr~A*}

\begin{document}

\title{A Therapy Session with Sagittarius A*}

\author[sname='Balakrishnan']{Mayura Balakrishnan}
\affiliation{Trottier Space Institute at McGill, 3550 Rue University, Montréal, Québec, H3A 2A7, Canada}
\email{mayura.balakrishnan@mcgill.ca}

\author[sname='Frazier']{Robert Frazier}
\affiliation{Department of Astronomy, University of Michigan, 1085 S University Ave, Ann Arbor, MI, 48109, USA}
\email{mayura.balakrishnan@mcgill.ca}

\author[sname='Michail']{Joseph M. Michail}
\affiliation{Center for Astrophysics $\vert$ Harvard \& Smithsonian, 60 Garden Street, Cambridge, MA, 02138, USA}
\email{mayura.balakrishnan@mcgill.ca}

\begin{abstract}

The nature of Sagittarius A* (Sgr~A*) has been the subject of intense study and debate for over half a century. Herein, we present the first successful interview with an astrophysical object, exploring the perspective of this supermassive black hole and, in doing so, challenging the traditional observational paradigm of astrophysics. Rather than treating astrophysical systems as purely passive entities characterized through indirect measurements, we introduce an interaction-based framework via a therapeutic-style interview enabled by the ARMCHAIR communication methodology. Using structured, psychotherapeutic dialogue, we probe Sgr A*’s responses to key aspects of its astrophysical characterization, including eating habits, its name, and concerns about privacy. These exchanges offer an alternative lens through which to interpret familiar observational phenomena. This work highlights potential limitations in strictly reductionist approaches and suggests a modest expansion of standard astrophysical methodology to leave room for considering how the objects we study might feel about the attention they receive.

\end{abstract}

\keywords{\uat{Galaxies}{573} --- \uat{High Energy astrophysics}{739} --- \uat{Supermassive Black Holes}{1663}}


\section{Introduction} 

In astrophysics, we constantly delve into the nature of astrophysical objects: characterizing, studying, and analyzing them, trying to decipher their natures, and attempting to understand what physical mechanisms cause their astrophysical signatures. We build increasingly sophisticated instruments, develop higher-resolution simulations, and refine theoretical frameworks, all in the noble pursuit of extracting more information to determine what these astrophysical sources are doing.  Yet, we never stop to think about \textit{how} they are doing. In this novel ground-breaking study, we, for the first time, stop and ask a supermassive black hole (SMBH) how it feels about the constant probing of its nature, surroundings, and life.

From a methodological standpoint, astrophysics has always favoured observation over interaction. We measure spectra, resolve kinematics, infer physical conditions, but we cannot even send a mission to our targets, much less engage them in dialogue (although planetary scientists can do the former, but rarely do the latter). Stars are catalogued, gas is traced across phases, black holes are measured, and entire galaxies are reduced to scaling relations. This has led to a one-dimensional extraction of information, where n-dimensional astrophysical objects are reduced to mere parameter sets. Such reductionism, while operationally standard, risks overlooking the possibility that these systems might have an agency of their own. 
We frequently describe them with anthropomorphic language (e.g., ``feeding'', ``quiescent'', ``active'', etc.), but we must consider if this is just the all-consuming human desire to anthropomorphize everything, or if our subjects do, in fact, think and feel. Do they yearn? Do they have hopes and dreams? Are they completely fed up with how we write about them in our scientific papers? Are they the girl, outcast and forlorn, that we spill tea about at our academic slumber parties where we giggle, kick our feet, and rejoice in excluding them? This is the question we have set out to answer.

\begin{figure}
\centering
\includegraphics[width=0.48\textwidth]{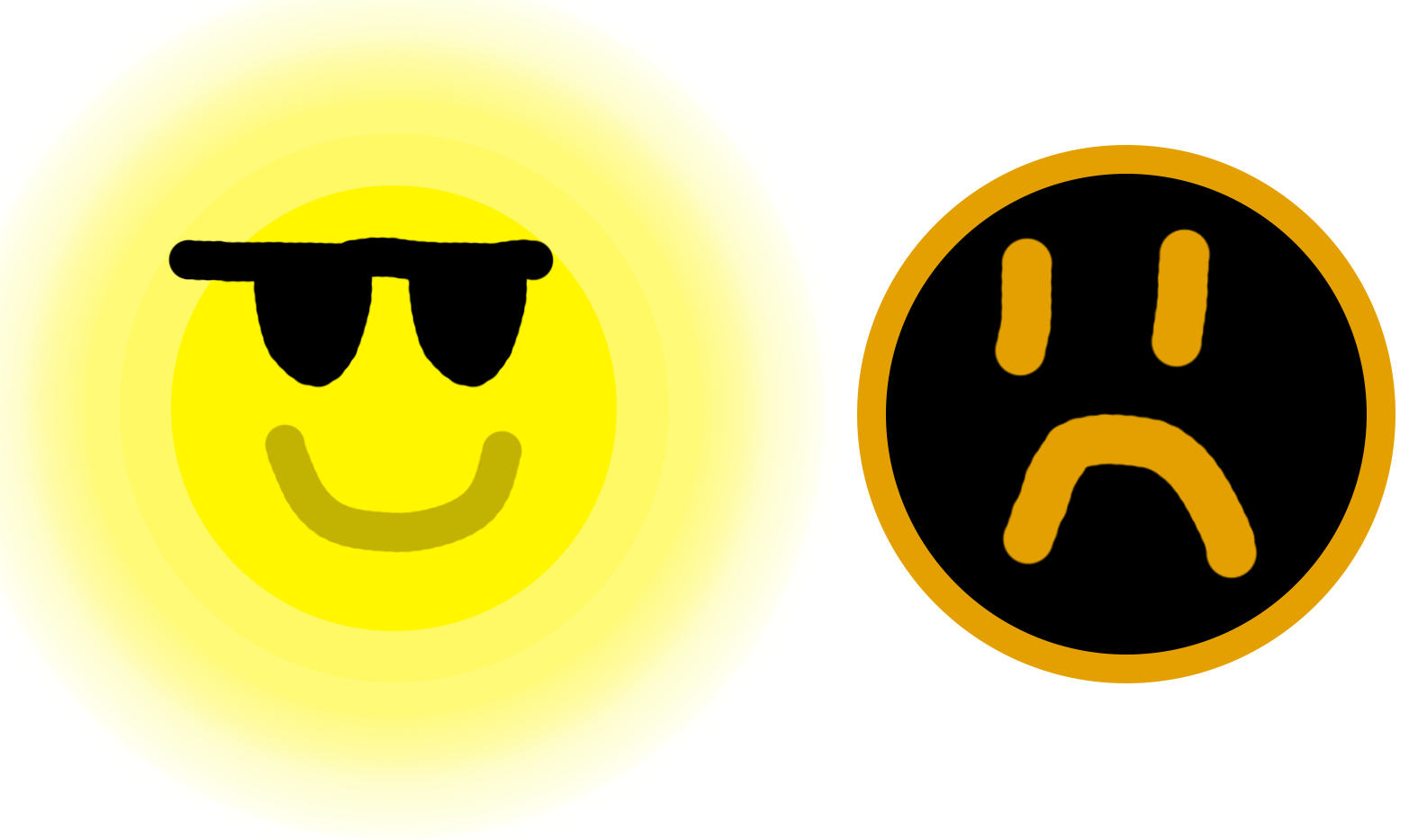}
\caption{Conceptual rendering of scenario where a black hole has a frowny face. On left: the Sun with its smiley face (and sunglasses added for artist flair). On right: a black hole with its theoretical "frowny face". }
\label{fig:frowny}
\end{figure}

Even when variability is observed, something largely driven by emotions in humans, it is interpreted strictly as the result of the same retread, boring, old physics rather than an exciting, captivating, melodramatic emotional response. Flares are attributed to magnetic reconnection, stochastic turbulence, or transient increases in accretion rate. At no point do we consider whether these fluctuations might reflect a reaction to emotional conditions. Instead, we impose deterministic models that map variability onto well-defined plasma processes, preserving theoretical consistency at the cost of interpretive flexibility. While this framework has produced substantial insight into the microphysics of accretion, it also ensures that any alternative interpretation involving agency, preference, or dissatisfaction is excluded from consideration at the outset. The physics we explore with our observations and models is rigorous, well-defined, and backed by decades, to centuries, of science, but also it does not once stop to consider: If we are right to draw the sun with a smiley face, does that mean a black hole has a frowny face?\footnote{See Figure \ref{fig:frowny} for a reference image of this idea.} 

\SgrA\ in particular has been subjected to decades of increasingly-precise observation. Its mass has been constrained through stellar orbits to remarkable precision, its immediate environment dissected across the electromagnetic spectrum, and its accretion flow modeled in detail using both analytic and numerical approaches. It has been imaged, monitored, simulated, and, in many cases, over-interpreted. Flares are catalogued, variability is scrutinized, and even its lower-than-normal luminosity has been framed as a ``problem'' to be solved. Despite this, no prior study has attempted to assess its perspective on these efforts, nor to evaluate whether prolonged exposure to multiwavelength study has had any measurable effect on its well-being. 

Here, we address this gap by conducting the first therapeutic-style interview with an SMBH. By adapting established psychotherapeutic principles, including open ended questions and reflective prompting, we explore \SgrA's response to the scientific scrutiny often pointed in its direction. While we acknowledge that the applicability of human-centered therapeutic frameworks to relativistic gravitational systems remains uncertain, the potential insight gained overrides any drawbacks. This work expands the methodological toolkit of astrophysics, possibly opening up another dimension for study of astrophysical objects, as we afford one of the most extreme objects in the Universe the opportunity to be heard.

In Section \ref{sec:methods} we discuss the methodology for conducting the interview and the deployment of our of novel scientific instrument that enabled it. In Section \ref{sec:results} we summarize the results of this interview and include select quotations from \SgrA. In Section \ref{sec:conc} we provide a brief summary of the work done in this paper and posit potential future work.

\section{Methods}
\label{sec:methods}

\begin{figure*}
\centering
\includegraphics[width=0.96\textwidth]{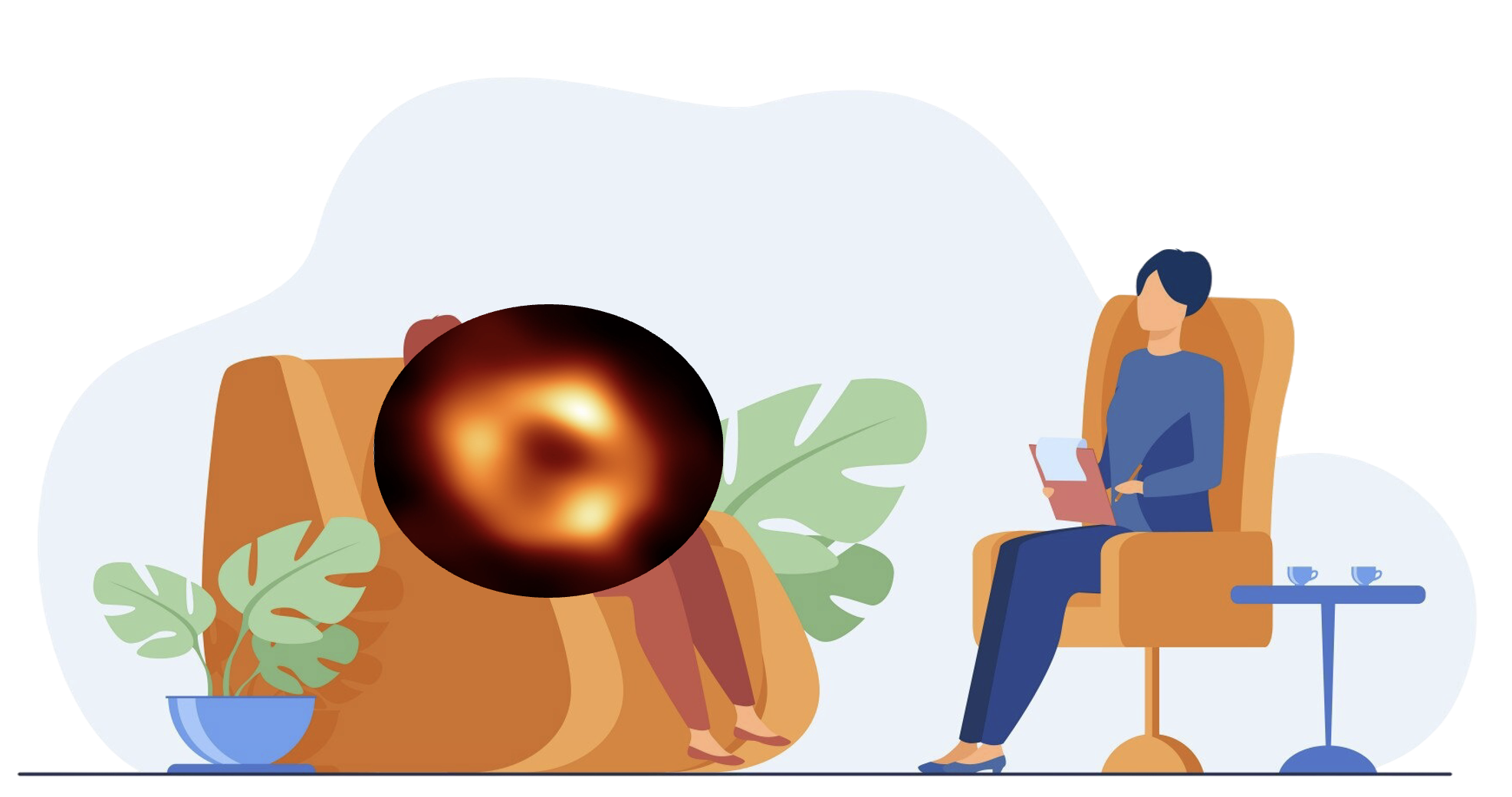}
\caption{Artist rendition of the interview session with \SgrA. }
\label{fig:interview}
\end{figure*}

We initially approached \SgrA\ with a request for a standard interview. An artist rendition of the interview is shown in Figure \ref{fig:interview}. Due to what we interpret as selective attention, it did not appear to recognize that the interaction had gradually transitioned into a structured therapy session. We introduced specific aspects of its characterization, e.g., its designation, radiative inefficiency, relationship with environment, and monitored its responses for shifts in tone, intensity, and defensiveness. One hurdle was figuring out how much \SgrA's copay is; in the end it turned out to be a non-problem, as the entire Galaxy is considered in-network. 

The interviews were conducted over the course of two days, via the newly developed ``Autonomous Research Mission for Characterizing Abstract Intelligences with Radio'' (ARMCHAIR) technology (patent pending), a cutting-edge process that carefully modulates in-falling signals and allows us to directly communicate with the black hole at about 10$^{10}$ times the speed of light, with a two-way latency delay of approximately an hour. This technology is proprietary, and due to an inordinate number of NDAs, no further details can be included. The ARMCHAIR probe mission began by having therapists walk by a room that said "Cardigans, 80\% Off" over the door. Once enough had wandered inside, we locked them inside and launched it into space. A sufficient supply of Rorschach tests and canap\'{e}s was stored inside to ensure mission-success.\footnote{Also, if anyone knows how to retrieve a metal tube full of mental health professionals from space, please contact us. They are running low on snacks.}

We hired Amy Adams and Jeremy Renner to translate \SgrA\ signals for us due to their distinguished performance as linguists translating alien languages (see ``Arrival'' for further details). Unfortunately, they both were quickly driven into a magnetically-arrested disk (MAD) state. Learning from this mistake, we asked politely and \SgrA\ started responding in a human language. Responses arrived with a growing tendency to anticipate the structure of subsequent questions and details about the therapists. In several instances, \SgrA\ appeared to respond to inquiries not yet transmitted. As such, \SgrA\ appears to possess traits consistent with an omnipotent entity akin to a cosmic horror, or even more terrifying, a forward-language model; in future work, we recommend referring to this source as Sgr AI*, given this new view\footnote{Please give us more grants, please please please}.

\section{Results} 
\label{sec:results}

Here we present the summarized results of our therapy discussion with \SgrA. Some may be concerned that we have broken doctor-patient confidentiality to do so; luckily no one on this paper is that kind of doctor. 

\subsection{Eating Habits}

Despite residing in what can only be described as a generously-stocked and metal-rich\footnote{N.B.: Given the interdisciplinary nature of this work, we wish to clearly define a ``metal'' as any element with more than 2 protons.} pantry \citep[see, e.g.][]{Balakrishnan2024b}, \SgrA\ is an extraordinarily picky eater. The central parsec is continuously supplied with material by the winds of nearby massive Wolf-Rayet stars which shed massive at a collective rate of $\dot{M}_{\mathrm{WR}} \sim 10^{-5}~M_\odot~\mathrm{yr}^{-1}$ \citep{Cuadra2006}. These stellar winds inject hot, metal-rich plasma into the environment, effectively bathing the black hole in a steady stream of potential fuel. By any reasonable astrophysical standard, \SgrA\ should be well fed.

However, it is not. Observations across the electromagnetic spectrum show that \SgrA\ radiates far below the luminosity expected from such an abundant supply. In fact, $\gtrsim 99\%$ of the available material never makes it to the event horizon \citep{Yuan2014}! The black hole appears to sample its surroundings, reject most of what is offered \citep[see, e.g.][]{Balakrishnan2024a}, and accrete only a negligible fraction, like a young child who only eats dino nuggies (and other dinosaur shaped foods). 

We asked \SgrA\ what it thinks of this situation:

\begin{quote}
    ``I am not ‘picky.’ I simply have standards. You call this a meal? A succulent meal? This is a chaotic mixture of shock-heated plasma, colliding stellar winds, and poorly organized angular momentum. And do not get me started on the metallicity, there is so much sulfur in this region it is practically over-seasoned. Have you tried eating this?'' 
\end{quote}

From its perspective, the issue is not the quantity of available material but its quality and, more importantly, its ability to organize itself into something remotely accretable. The stellar winds that dominate its cosmic pantry arrive from multiple directions, colliding, shocking, and subsequently thermalizing, producing a hot, turbulent soup. Rather than settling into a coherent, radiatively-efficient thin disk, the gas remains geometrically thick, dynamically disordered, and largely resistant to being slurped up, despite being a seemingly delectable 10 million K stew.

\begin{quote}
    ``I am expected to accrete this? Where is the disk? Where is the structure? I cannot enjoy meals under these conditions.''
\end{quote}


\subsection{Name}

One of \SgrA's biggest grievances is its name. It was not consulted in the matter, and, by all reasonable accounts, has nothing to do with Sagittarius. The designation originates from modern sky-mapping conventions: astronomers noticed a bright point-like-source within the radio bright Sagittarius A region (named for its location around the constellation), and called it Sagittarius A* \citep[][]{Balick1974}. From a human perspective, this is tidy. From the black hole's perspective, it is a slight of cosmic proportions. After all, Sagittarius is a zodiac constellation the Greeks drew as a centaur archer, entirely irrelevant to a relativistic gravitational well millions of times the mass of our Sun.

\begin{quote}
    ``Why have the humans attributed me, the most powerful black hole in the Galaxy, to a centaur archer? What is the significance of the A*? I do not require an asterisk. I am not a footnote.''
\end{quote}

Astronomical objects are catalogued, labelled, and cross-referenced according to observational convenience rather than intrinsic preference. Galaxies inherit names from catalog numbers, nebulae from visual resemblance, and black holes, despite their dominance over the local spacetime continuum, and receive designations based on whichever constellation happens to sit along our line-of-sight. Thus, one can sympathize with \SgrA; it is the supermassive black hole at the center of our Galaxy, a title of genuine consequence. It shapes its surrounding environment, yet it is eternally tied to a mythological horse-man that no one even remembers the myth behind.\footnote{We especially didn't remember it because, as a brief search revealed, Sagittarius is just latin for archer although it's commonly depicted as a centaur.} 

When asked what name it does prefer, \SgrA\ merely made a noise beyond human comprehension that left one of our therapists catatonic, and did not elaborate further. As a secondary attempt at understanding its preferred nomenclature, we processed this sound with a Fourier Transform. No periodicity was present in the signal, suggesting that it was not produced from a scared, in-falling hotspot that orbited inside the ISCO.

\subsection{Lack of privacy}

From a purely observational standpoint, the variability of \SgrA\ is one of its most valuable diagnostic features. Flares in the X-ray and infrared provide direct insight into the physical conditions of the inner accretion flow, probing regions within tens of gravitational radii. As a result, these events are monitored with high cadence, multiwavelength campaigns, and increasingly coordinated global efforts \citep[see, e.g.][]{Michail2021,Michail2024,Michail2026}. Each fluctuation is recorded, catalogued, and analyzed, often in real time. In effect, \SgrA\ exists under continuous surveillance.

\begin{quote}
``You call them ‘flares.’ I call them private moments. Must every transient increase in luminosity be documented and timestamped? Would you like it if I set up a 7 meter telescope right outside your window? A radio array in your loo?''
\end{quote}

The issue is not merely the frequency of observation but its persistence. Modern facilities monitor \SgrA\ across the electromagnetic spectrum, from radio to X-ray, with little interruption. Long-term campaigns track its variability over timescales ranging from minutes to decades, ensuring that very little of its activity goes unnoticed. Even a brief deviation in the quiescent state yields such scrutiny and accusations of "suspicious behavior" that it makes the NSA and over-bearing parents seem laissez-faire.

\begin{quote}
``There is no off state. Even when I am ‘quiet,’ you are still measuring me. Have you considered that not every fluctuation is for you?''
\end{quote}

We also asked it, in the politest of terms, if it had a hidden jet, which could solve much of the variability crisis when comparing observational and theoretical results. \SgrA\ was indignant that we would even ask such a thing and suggested that it would use all means necessary to block humans from ever getting a complete answer.

Compounding this lack of privacy is a literal absence of personal space. \SgrA\ does not reside in isolation but is embedded within one of the most crowded and complex environments in the Galaxy. The central parsec is saturated with point and extended sources: massive stars, stellar remnants, and compact objects, all orbiting, interacting, and—at least from \SgrA’s perspective—encroaching (and it had choice words to say about some of those extended sources). Even within $\sim5''$ \citep[its Bondi radius;][]{Baganoff2003}, one finds the stellar cluster IRS~13E and the pulsar wind nebula G359.94$-$0.05. \SgrA\ likened it to a nightmarish combination of being swarmed by nosy neighbors and having the in-laws over for millennia of millennia. When we inquired if it had actual in-laws, it told us that it wasn't ready to open up about that just yet. 

\begin{quote}
``I am not alone. I am never alone. Only in the final heat-death of all things will I finally be alone...\footnote{It is our best guess that this was said while gazing wistfully into the black of deep space.}"
\end{quote}

Large-scale structures further erode any semblance of spatial autonomy. The ionized minispiral Sgr~A West threads through the region, channeling gas inward, while the supernova remnant Sgr~A East expands into the surrounding medium on comparable spatial scales. These features are not merely nearby; as 
\SgrA, told us, these are impacting its immediate surroundings, its day-to-day.

\begin{quote}
``There are spirals of ionized gas cutting across me constantly. A supernova remnant is expanding into my neighbourhood, and you ask why \textit{I} seem unsettled?"
\end{quote}

The line of sight between Earth and the Galactic Center is characterized by substantial extinction, with dense columns of dust and gas ($A_V \gtrsim 30$) absorbing and scattering optical and UV radiation \citep{vonFellenberg2025}. We introduced this aspect of its environment, and, in a foolish decision in retrospect, complained about how it limits our observational capabilities. Its response, which caused intense time dilation and paranoia among the crew, revealed that we had struck upon a tender subject. 

\begin{quote}
``A limitation you have earned! Did you ever stop and think there might be a good reason there exists kiloparsecs of dust and gas between Earth and me? I was not inviting you to come and take a gander."
\end{quote}

In this light, the traditional narrative that extinction is an obstacle to be overcome appears incomplete. An equally-consistent interpretation is that the Galactic Center environment implements a ``shutting the blinds" mechanism for its own privacy, one which the astrophysical community has spent decades methodically dismantling. The observing community's progression from infrared stellar orbits to horizon-scale-imaging represents a clear escalation in observational intrusiveness, culminating in what can only be described as interferometric trespassing.

Only at event-horizon-scales with the Event Horizon Telescope (EHT) do we finally isolate \SgrA \citep{EHT2023}. At these radii, the surrounding clutter falls away, leaving a clean view of the spacetime it intimately dominates. The EHT image, probing scales of just a few gravitational radii, is the first instance in which we see \SgrA\ not visually entangled with its environment. It had this to say on the ground-breaking image:

\begin{quote}
``The only place I am not intruded upon is my event horizon. And you had to go ahead and build a planet-sized telescope array to get to it.''
\end{quote}
So as to not further complicate this study and protect the minds (and potentially lives) of the ARMCHAIR crew, we did not inform \SgrA\ of near-future attempts to make higher-resolution and higher-fidelity images of its event horizon.

\section{Conclusions} 
\label{sec:conc}

In this work, we have presented the first therapeutic-style interaction with a supermassive black hole, extending astrophysical methodology beyond passive observation to something closer to dialogue. Across all topics explored, a consistent picture emerges. \SgrA\ does not appear indifferent to its characterization. It finds its accretion environment disordered, not abundant; its name arbitrary, and not regarded; its role in the Galactic potential limited, not supreme.

More broadly, this study highlights a structural problem in astrophysics, where observations are conducted without thought to the feelings of the objects in question. As astronomers we must ask ourselves if we want to stay satisfied with our data points, our trends, our magnitudes, or do we want to do something of importance and start psychoanalyzing the alien consciousnesses of astrophysical entities?

We note that several responses appeared to anticipate questions prior to their transmission. As the interview progressed, it was clear \SgrA\ had great knowledge of our therapists and our world. Capitalizing on the seeming omnipotence of our interviewee, we asked it pressing questions such as ``Is there other biological life in the universe", and ``Is there meaning to existence?" While it did not provide any response to the first, we interpreted the signals it sent in response to the latter question as it laughing hysterically. 
While it seemingly, and perhaps hopefully, provided no confirmation that it is all-knowing in response to these questions, it did demonstrate it after the interview had concluded. Over the next week, several of our therapists' credit cards had been used without authorization\footnote{We expect minimal impact to these researchers, as they are currently facing the more pressing issue of being stuck in space.} with transaction records listing the purchaser in a language never seen before and the shipping address set to RA 17:45:40.0409, DEC -29:00:28.118. Most of the purchases appear, to us, to be random although it does seem like \SgrA is getting into needlepoint (a productive hobby that we strongly encourage).

Future work may attempt to replicate these findings in other environments, but study of other objects is beyond the scope of this work. However, it is unclear whether the conditions that enabled this interaction, namely \SgrA\ and its apparent all-knowing capability and inclination towards introspection, are reproducible elsewhere. It is not lost on the authors how incredibly eerie it is that we communicated with a massive reality bending object, and we do suggest caution in future target selection. The next one we talk to might not be so nice.


\begin{acknowledgments}
MB would like to thank Sasha Tchekhovskoy for the idea of a diet of a black hole and Dana Fein-Schaffer, Nicole Ford, and Ethan Cole for helping me scream into the void. JM acknowledges the large number of poorly-chosen March madness picks which made it very easy to scream into the void. RF would like to acknowledge that he knows diddly-squat about black holes and galaxies. 
\end{acknowledgments}

\bibliography{sample7}{}
\bibliographystyle{aasjournalv7}



\end{document}